\documentclass[aps,prb,twocolumn,showpacs,superscriptaddress,floatfix,nofootinbib,nobibnotes]{revtex4-1}
\usepackage{amsfonts}
\usepackage{amsmath}
\usepackage{amssymb}
\usepackage{graphicx}
\usepackage{float}
\usepackage{mathtools}

\DeclareFontEncoding{LGR}{}{}
\DeclareTextSymbol{\~}{LGR}{126}
\newcommand{\lyxmathsym}[1]{\ifmmode\begingroup\def\b@ld{bold}
 \text{\ifx\math@version\b@ld\bfseries\fi#1}\endgroup\else#1\fi}

\begin{document}

\title{Charge carrier transport asymmetry in monolayer graphene}

\date{\today }

\author{E. Zion}

\affiliation{Institute of Nanotechnology and Advanced Materials, Bar-Ilan University,
Ramat Gan 52900, Israel }

\author{A. Butenko}

\affiliation{Institute of Nanotechnology and Advanced Materials, Bar-Ilan University,
Ramat Gan 52900, Israel }

\author{A. Sharoni}

\affiliation{Institute of Nanotechnology and Advanced Materials, Bar-Ilan University,
Ramat Gan 52900, Israel }

\author{M. Kaveh}

\affiliation{Jack and Pearl Resnick Institute, Department of Physics, Bar-Ilan
University, Ramat Gan 52900, Israel }

\author{I. Shlimak}

\email[Corresponding author: ]{ishlimak@gmail.com}
\affiliation{Jack and Pearl Resnick Institute, Department of Physics, Bar-Ilan
University, Ramat Gan 52900, Israel }

\begin{abstract}

The conductivity and Hall effect were measured in CVD-grown monolayer graphene as a function of the gate voltage, $V_{g}$, at temperatures down to $T$ = 2 K and in magnetic fields up to $B$ = 8 T. The minimal conductivity was observed at positive $V_{g}$ which shows the position of the charge neutrality point, $V_{NP}$. With decreasing $T$, $V_{NP}$ first decreases, but stop to decrease at low $T$. Hysteresis of conductivity shows similar behavior: it decreases with decreasing $T$ and disappears at low $T$. A significant asymmetry was observed at low density of charge carriers $|n|=(n,p)$: mobility of holes was less than mobility of electrons. The asymmetry decreases with increasing $|n|$. It was observed that the value of $|n|$ determined from the Hall effect is less than the full value induced by $V_{g}$. In strong perpendicular $B$, Shubnikov-de Haas (SdH) oscillations were observed in the longitudinal conductivity, $\sigma_{xx}$, together with half-integer quantum Hall plateaus. It was found that $|n|$ determined from SdH oscillations is equal to the full value induced by $V_{g}$ as opposed to the Hall effect. Explanatory models for all observed phenomena are discussed. 

\end{abstract}

\pacs{73.22.Pr}

\maketitle

\section{introduction}

There are plenty of experimental and theoretical papers devoted to the investigation of the charge carrier transport in monolayer graphene, including measurements in the field-effect-transistor (FET) configuration, where the density of charge carriers, $|n|$ (electrons $n$ and holes $p$), and corresponding position of the Fermi level, $E_{F}$, can be controlled by varying the gate voltage $V_{g}$ (see \cite{1,2,3,4} and references therein).\\

In ideal neutral graphene, $E_{F}$ lies at the Dirac point, $E_{D}$, which separates the region of conduction by electrons (when $E_{F}>E_{D}$ and lies within the conduction band) from the region where transport is governed by holes (when $E_{F}<E_{D}$ and lies within the valence band). Because this is a single point in the band structure, the density-of-states is zero. Therefore there are no states to occupy and hence no carriers (at the Dirac point the carrier concentration vanishes, $n_{D} = p_{D} = 0$). At thermal equilibrium, the position of $E_{F}$ is constant across the sample. However, in real samples, the position of $E_{D}$ with respect to the Fermi level, $\Delta E_{F}=|E_{F}-E_{D}|$, fluctuates locally due to corrugations and non-flatness of the monolayer graphene film. Even more significant is the shift of $\Delta E_{F}$ due to chemical potential variations induced by charged impurities in the substrate, in the vicinity of the graphene layer. Charge impurities and charge adsorbates modify electrostatic potential and lead to the local fluctuations of $\Delta E_{F}$. As a result, the electrostatic gating is not homogeneous within the graphene sheet and at the Dirac point, the system splits into hole-rich and electron-rich puddles \cite{5}. This gives rise to the finite and equal average carrier density at the Dirac point ($<n_{D}> = <p_{D}> \neq 0$) which is therefore called as the "charge neutrality point" (NP). At NP, the Hall resistance,$R_{xy}$, and Hall coefficient, $R_{H} = R_{xy}/B$, crosses zero due to the equality $<n_{D}> = <p_{D}>$, and conductivity reaches its minimal value, $\sigma_{min}$. Usually $\sigma_{min} \approx 4e^{2}/h$, but other values were also observed \cite{6} ($e^{2}/h = 38.7$ $\mu$S is the quantum of two-dimensional conductivity).\\

It was found in many previous investigations that for graphene films supported by different substrates, the position of the gate voltage at the NP, $V_{NP}$, is located at different negative or positive values. This is explained by unintentional doping - existence of some amount of positively or negatively charged impurities (called respectively "donors" or "acceptors") which produce corresponding concentration of charge carriers. The screened potential of these charged impurities is also responsible for scattering of charged carriers which limits the mobility. Mechanisms of scattering by charge impurities, as well as scattering on neutral centers like ripples, vacancies, grain boundaries, etc. are reviewed in\cite{7}.\\

The final aim of this work is the investigation of the influence of ion irradiation on the electron transport in a large scale (5x5 mm) CVD grown monolayer graphene samples offered on the market by $Graphenea$ company. In view of the above, any modification of graphene needs preliminary careful measurements of samples in the initial state which can be quite individual. In this work, we report features of electron transport in initial samples before irradiation. Variation of these properties after irradiation will be the subject of further work.

\section{samples}

According to certificate, polycrystalline monolayer graphene films were grown by CVD method on copper catalyst and then were transferred to SiO$_{2}/$Si substrate, the width of insulating SiO$_{2}$ layer was 300 nm ($\pm 5\%$), $p$-Si substrate was heavily doped, with resistivity $\rho < 0.005$  $\Omega cm$. In our previous work \cite{8,9,10,11,12}, the results were reported concerning investigation of the Raman scattering spectra and two-probe conductivity in two series of micro-size samples made on the surface of the large-scale specimen. These samples were subjected to irradiation by different ions, followed by ageing and annealing of the radiation damage.\par

For the present work, a new similar large-scale specimen supplied by $Graphenea$ company was patterned via photolithography followed by O$_{2}$ milling into the Hall-bar shape. The width of the graphene strip in each micro-sample was $W = 10$ $\mu$m, the total length between source and drain was $100$ $\mu$m and the length between voltage probes was $L = 30$ $\mu$m. Metallic contacts were deposited using e-beam evaporation at room temperature, in a high vacuum chamber (base pressure $\sim 10^{-8}$ Torr). Contacts were made of 5 nm Ti and 80 nm Au. The low resistive Si substrate was used as the gate electrode (Fig. 1). Measurements were performed in a commercial Quantum Design PPMS-9 cryostat, via a 4-probe configuration in the temperature interval from 300 K to 2 K, and in magnetic fields $B$ up to 8 T. The values of longitudinal, $V_{xx}$, and transversal, $V_{xy}$, voltages were  measured at constant current, $I=10$ $\mu$A. 

The measurements of conductivity $\sigma_{xx} = 1/\rho_{xx} = (L/W)(I/V_{xx})$ at 300 K preformed just after sample preparation, showed that $\sigma_{xx}$ continuously decreases when sweeping $V_{g}$ from – 80V to + 80V. This means that graphene was heavily doped with acceptors like polymer traces and adsorbed molecules of water and oxygen from the ambient air. In order to remove the polymer traces, the samples were first annealed in a cylindrical furnace, under a constant flow (1000 sccm) of a mixed forming gas: 95$\%$ Ar + 5$\%$ H$_{2}$ at 250$^{\circ}$C for 1h \cite{12}. Most of the adsorbed molecules were removed by a second annealing $in situ$ when samples were already mounted into the cryostat. In this annealing, all samples were subjected to a long-time (17 hrs) heating at 150$^{\circ}$C in a vacuum of 10 Torr. The long-time annealing allows one to observe the minimum of $\sigma_{xx}$ which indicates the  charge neutrality point (NP). The exact position of NP was determined by the value of $V_{g}$ where $R_{xy}$ crosses zero. The results obtained from one of the samples are presented below. Measurements performed on other samples showed similar results which evidences their reliability. 

\begin{figure}[H]
	\centering
	\includegraphics[scale=0.29]{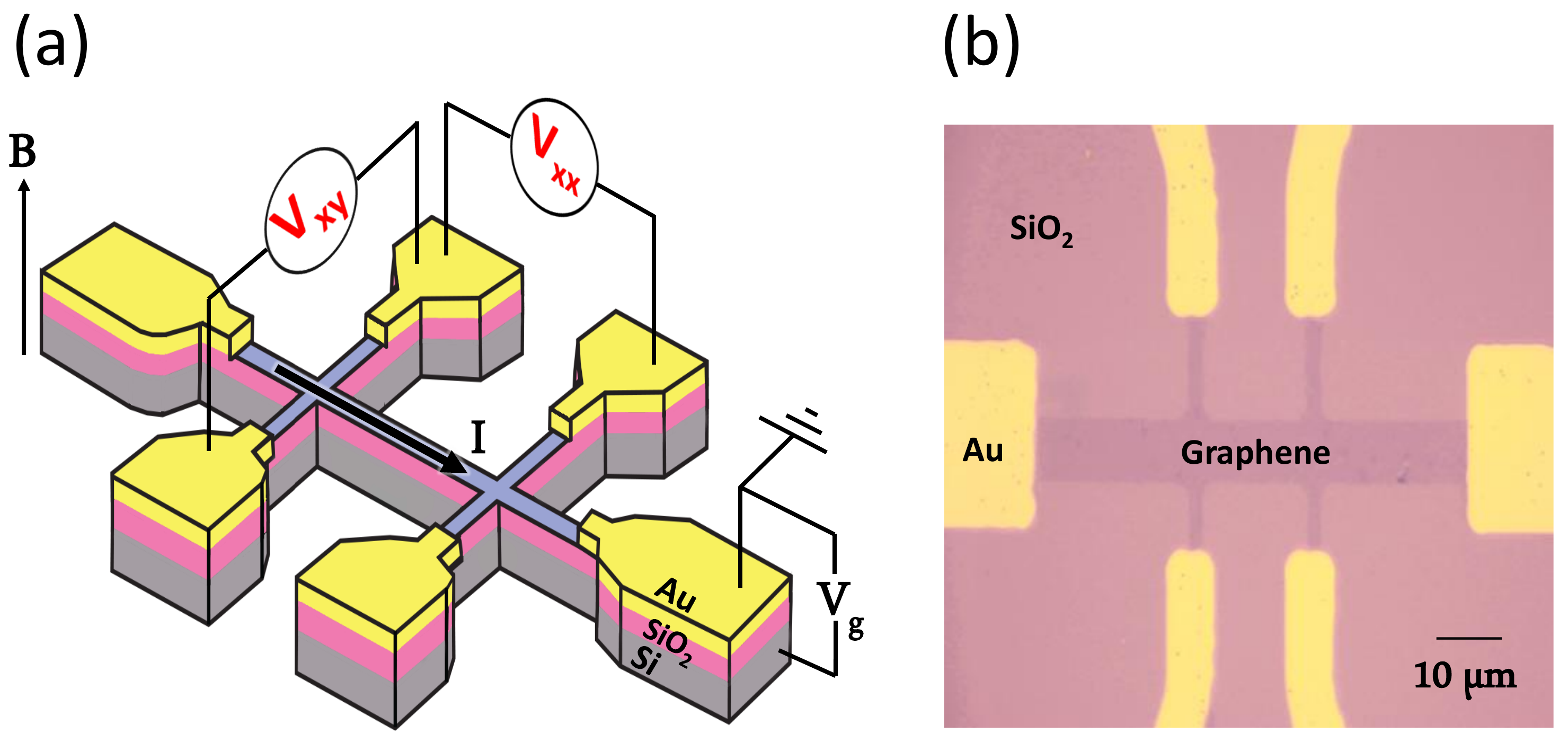}
	\caption{(Color online) (a) Schematic
illustration of sample structure and measurement geometry. (b) Optical microscopy image of a single sample, graphene is visible as a blue strip.}
	\label{Layout1}
\end{figure}

\section{results and discussion}

a) \underline{Conductivity.}\\

Figure 2 shows the conductivity $\sigma_{xx}$ plotted as a function of $V_{g}$ at different temperatures. The solid and dashed lines correspond to scanning $V_{g}$ from -80 V to +80 V and back. Besides the asymmetry of conductivity, which will be discussed later together with asymmetry of mobility, the following three features can be seen: (i) $V_{NP}$ is located at positive voltage at all $T$; with decreasing $T$, $V_{NP}$ first shifts to lower $V_{g}$ and then saturates (see insert in Fig. 2); (ii) hysteresis is observed at high $T$ (300 and 200 K), but disappears below 100 K together with saturation of $V_{NP}$; (iii) reproducible conductivity fluctuations are observed at large $V_{g}$ and low $T$.  We now discuss these various features.\par

\begin{figure}[H]
	\centering
	\includegraphics[scale=0.38]{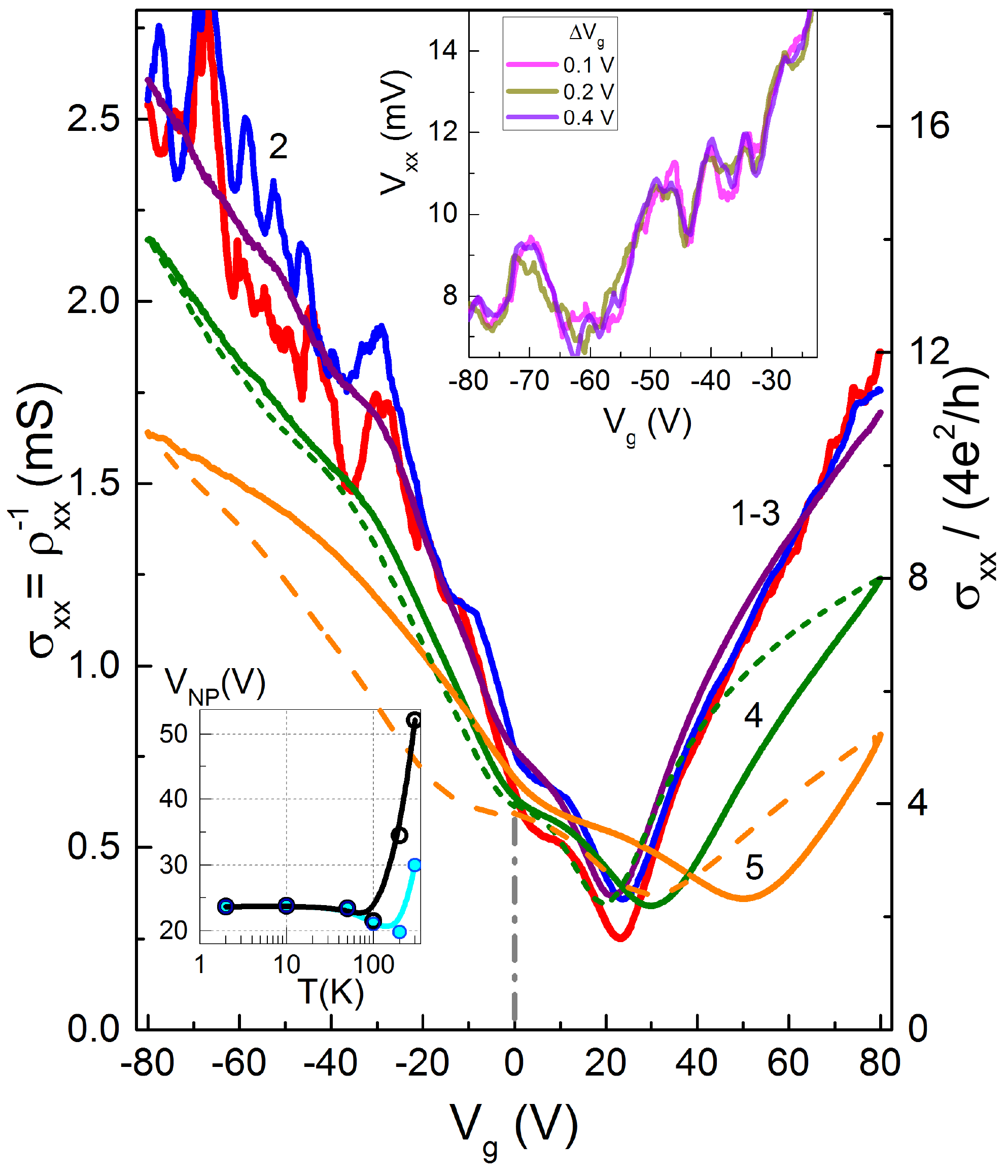}
	\caption{(Color online) Longitudinal conductivity $\sigma_{xx} = (\rho_{xx})^{-1}$ as a function of the gate voltage $V_{g}$ at different temperatures $T$. Conductivity is shown in mS (left scale) and in dimensionless units $\sigma/(4e^{2}/h)$ (right scale). The solid and dashed lines correspond to scanning $V_{g}$ from -80 V to +80 V and back. Minimum conductivity corresponds to $V_{NP}$ - position of $V_{g}$ at charge neutrality point. $T$: 1 (red) – 2 K, 2 (blue) – 10 K, 3 (purple) – 100 K, 4(green) – 200 K, 5 (orange) - 300 K. Upper insert illustrates reproducibility of fluctuations: coincidence of $V_{xx}(V_{g})$ curves measured with different sweep rate $\Delta V_{g}$: 0.1, 0.2 and 0.4 V. Lower insert shows the shift of $V_{NP}$ with decreasing $T$. Two values of $V_{NP}$ reflect the width of hysteresis.}
	\label{Layout2}
\end{figure}

(i) Location of $V_{NP}$ at positive voltage shows the unintended doping of initial samples by holes due to existence of residual acceptor impurities. The shift of $V_{NP}$ to lower voltages can be explained by decreasing of holes due to freeze-out of holes on acceptors. However, it follows from Fig. 2 that at low $T$, $V_{NP}$ saturates at $V_{NP} \approx 20$ V. This means that the part of holes cannot be trapped by acceptors. A possible explanation is that some acceptors with density $N_{A}^{(S)}$ are located not on graphene, but in the SiO$_{2}$ substrate close to the graphene layer. The corresponding holes in graphene are, therefore, separated from acceptors by potential barriers which results in significant increase in the time of trapping of  holes by these acceptors at low temperatures during measurement time. One can estimate the concentration of these holes (and, correspondingly, $N_{A}^{(S)}$) from the compensating negative charge induced by the positive voltage $V_{NP}$: $N_{A}^{(S)} = (1/e) C_{ox} V_{NP}$. Here $C_{ox}$ is the capacitance of the insulating SiO$_{2}$ layer: $C_{ox} = \varepsilon_{0}\varepsilon/t$, where $\varepsilon_{0} = 8.85\times 10^{-12}$ F/m is the vacuum permittivity, $\varepsilon = 3.9$ is the dielectric constant of SiO$_{2}$ and $t = 300$ nm is the thickness of SiO$_{2}$ layer. Thus, we get $C_{ox} = 1.15\times 10^{-4}$ F/m$^{2}$ and for $V_{NP} \approx 20$ V, we obtain $N_{A}^{(S)} \approx 1.5\times 10^{12}$ cm$^{-2}$.\par

(ii) Hysteresis is usually explained by charging and recharging of impurity centers by the charge carriers. At fixed temperature, the width of hysteresis obviously depends on the sweeping rate \cite{13}: for slow sweeping, hysteresis decreases because there is more time for recharging. In our case, the decrease of the width and disappearance of hysteresis below 100 K shown in Fig. 2 can be explained by neutralization of part of acceptors due to freeze-out of holes and by significant increasing the effective time needed for recharging acceptors located in the SiO$_{2}$ substrate.\par

(iii) The conductivity fluctuations (CF) observed at low $T$ and large $V_{g}$ are highly reproducible. It is confirmed by repeated measurements at different sweep rates (see insert in Fig. 2). Reproducible CF were often observed in monolayer graphene (see, for example, \cite{14,15} and references therein) and explained by the long-range disorder potential induced by the randomly located charged impurities. This leads to a slow variation of the background potential as the Fermi energy is varied. As a result, the phase interference of charge carriers varies randomly which leads to CF. In our samples, CF were not observed by sweeping the magnetic field at fixed $V_{g}$. This can be explained by the smoothing of the conductance landscape induced by magnetic fields \cite{15}.\\

b) \underline{Concentration of charge carriers.}\\

Figure 3 shows concentration of charge carriers $n = 1/eR_{H}$ as a function of normalized gate voltage $V_{G} \equiv (V_{g} - V_{NP})$ measured at different $T$. Here $R_{H}$ is the Hall coefficient $R_{H} = R_{xy}/B$, the value $R_{xy} = V_{xy}/I$, was obtained after averaging of four measurements $V_{xy}$ at opposite directions of $I$ and $B$. It is generally assumed that concentration of induced carriers has to be linearly proportional to the gate voltage: $n = \alpha_{0}V_{G}$, where the slope $\alpha_{0} = (1/e)\cdot C_{ox}$, is determined by the gate capacitance.  In our case, with $C_{ox}=1.15 \times 10^{-4}$ F$/$m$^{2}$, $\alpha_{0} = 0.72\times 10^{11}$ cm$^{-2}$V$^{-1}$. This slope is shown as dashed line in Fig. 3. The slopes $\alpha$ are less than $\alpha_{0}$, they are different for electrons and holes, they increase as $T$ decreases from 300 to 100 K. Below $T = 100$ K, they are constant but smaller than $\alpha_{0}$. This behavior correlates with the temperature dependence of the width of hysteresis in conductivity shown in Fig. 2. Taking into account that hysteresis is caused by recharging of impurity centers, one can conclude that deficit of carrier concentration is also determined by the trapping of some carriers induced by $V_{G}$. In this case, the small difference between the measured electron and hole concentrations at equal $V_{G}$ is explained by the difference in concentration of trapped carriers. It was noted already that some charge carriers have to overcome potential barriers to reach the trapping centers located in the substrate. As a result, the probability of trapping decreases with decreasing $T$, and correspondingly, the concentration of carriers detected by the Hall-effect measurements increases. However, even at low $T$, the value of $\alpha$ is still less than $\alpha_{0}$. This hints that besides localization on the trapping centers, there is another mechanism which increases
the Hall constant, $R_{H}$, and correspondingly decreases the value of $|n| = 1/eR_{H}$. This is possible when charge carriers have different mobility. If, for example, full concentration of electrons $n$ consists of two sorts of carriers $n_{1}$ and $n_{2}$ with different mobility $\mu_{1}$ and $\mu_{2}$ , the expression for $R_{H}$ will be
\begin{center}
$R_{H} \approx \frac {(n_{1} \mu_{1}^{2}+n_{2} \mu_{2}^{2})}{e(n_{1} \mu_{1}+n_{2} \mu_{2})^{2}} \approx \frac {1}{en_{1}}[\frac {1+ab^{2}}{(1+ab)^{2}}]$
\end{center}
were $a = n_{2}/n_{1}$ , $b = \mu_{2}/ \mu_{1}$ . In the case when $ab>>1$, $|n| \approx an_{1} = n_{2}$ which is less than the full concentration $n$.
One can assume that charge carriers in our polycrystalline samples have different
mobility at different places (for example, near the grain boundaries), which leads to reduction of the concentration $|n|$ determined from the Hall effect measurements.\par

One can also see from Fig. 3 that straight lines $|n| = \alpha V_{G}$ from both sides do not intersect the ordinate axis at $V_{G} = 0$.  The cut-off values correspond approximately to $\pm 0.25\times 10^{12}$cm${-2}$. This can be attributed to the average concentration in the hole-rich $<p_{D}>$ and electron-rich $<n_{D}>$ puddles coexisting at NP. The value of $<n_{D}>$ was estimated in \cite{16} on the basis of the assumption that the spatial size of puddles is equal, while the random potential relief can be approximated by a step function of height $\pm \Delta$ (see insert in Fig. 3). In this model, when $\Delta >> kT$,
\begin{center}
 $<n_{D}>=\frac{2}{\pi\hbar^{2}v^{2}_{F}}(\frac{\Delta^{2}}{2}+\frac{\pi^{2}}{6}k^{2}_{B}T^{2})$ (Eq. 1)
\end{center}
Here $v_{F} = 10^{6}$ m/s is the Fermi velocity of the Dirac fermions. This gives
\begin{center}
$\Delta \approx \hbar v_{F} \sqrt[•]{\pi<n_D>}$ (Eq. 2)
\end{center}
In our case with $<n_{D}> = 0.25\times 10^{12}$cm$^{-2}$, we obtain $\Delta \approx 58$ meV, which agrees with the value $\Delta \approx 54$ meV determined for monolayer graphene in \cite{16} using the different method. \par

\begin{figure}[H]
	\centering
	\includegraphics[scale=0.38]{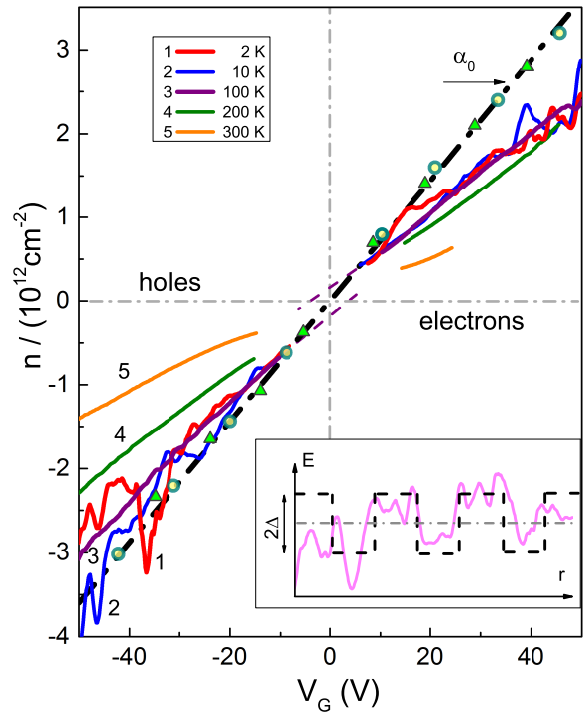}
	\caption{(Color online) Dependence of charge carrier concentrations as function of the reduced gate voltage $V_{G} \equiv (V_{g}-V_{NP})$ at different temperatures. $T$: 1 (red) – 2 K, 2 (blue) – 10 K, 3 (purple) – 100 K, 4(green) – 200 K, 5 (orange) - 300 K. The dot-dashed straight line corresponds to the slope $\alpha_{0} = 7.2\times 10^{10}$ cm$^{-2}/$V. The dashed lines show extrapolation of the average experimental dependences $|n|(V_{G})$ at low $T$ to the Y-axes. The open circles and triangles show the values of $n$ and $p$ determined from the SdH minimums in magnetic fields $B = 8$ T (open circles) and 7 T (triangles). Insert presents schematics of the step-like simplification of the random potential relief.}
	\label{Layout3}
\end{figure}

Fig. 3 shows also that at low $T$, concentration of charge carriers fluctuates as a function of $V_{G}$. These fluctuations, unlike the conductivity fluctuations, depend on time and are not reproducible.  They can be explained by the following considerations. It is known from investigation of the silicon-based FET \cite{17} that charged oxide defects with density $N_{t}$ inevitable occur in SiO$_{2}$ close (typically within 1-3 nm) to the conductive channel (monolayer graphene in our case). These defects can have different charge states and capable to be recharged by carriers from conductive channel. As a result, near-interfacial traps are sensitive to the Fermi level position: they tend to empty if their level $E_{t}$ is above the position of the Fermi level $E_{F}$ and capture electrons if their level is lower. Donor-like traps are positively charged in empty state and neutral in filled position (0/+), acceptor-like traps are negatively charged in a filled state and neutral while empty (-/0). Increase of $V_{g}$ is accompanied by the definite increase of $E_{F}$ at the interface which leads to changes in the charging state of these impurities. In its turn, this leads to the rise of fluctuations in both dependencies $\sigma_{xx}(V_{G})$ and $|n|(V_{G})$, which are more pronounced at low temperatures (Fig. 4). No correlation between fluctuations of dependencies $\sigma_{xx}(V_{G})$ and $|n|(V_{G})$ was observed. This shows that in spite of the fact that both fluctuations are induced by charged impurities, the physical mechanism is different: fluctuation of $|n|$ are caused by the occasional distribution of impurities in energy, while conductivity fluctuation originates from the random distribution of charged impurities in space.

 \begin{figure}[H]
 	\centering
	\includegraphics[scale=0.33]{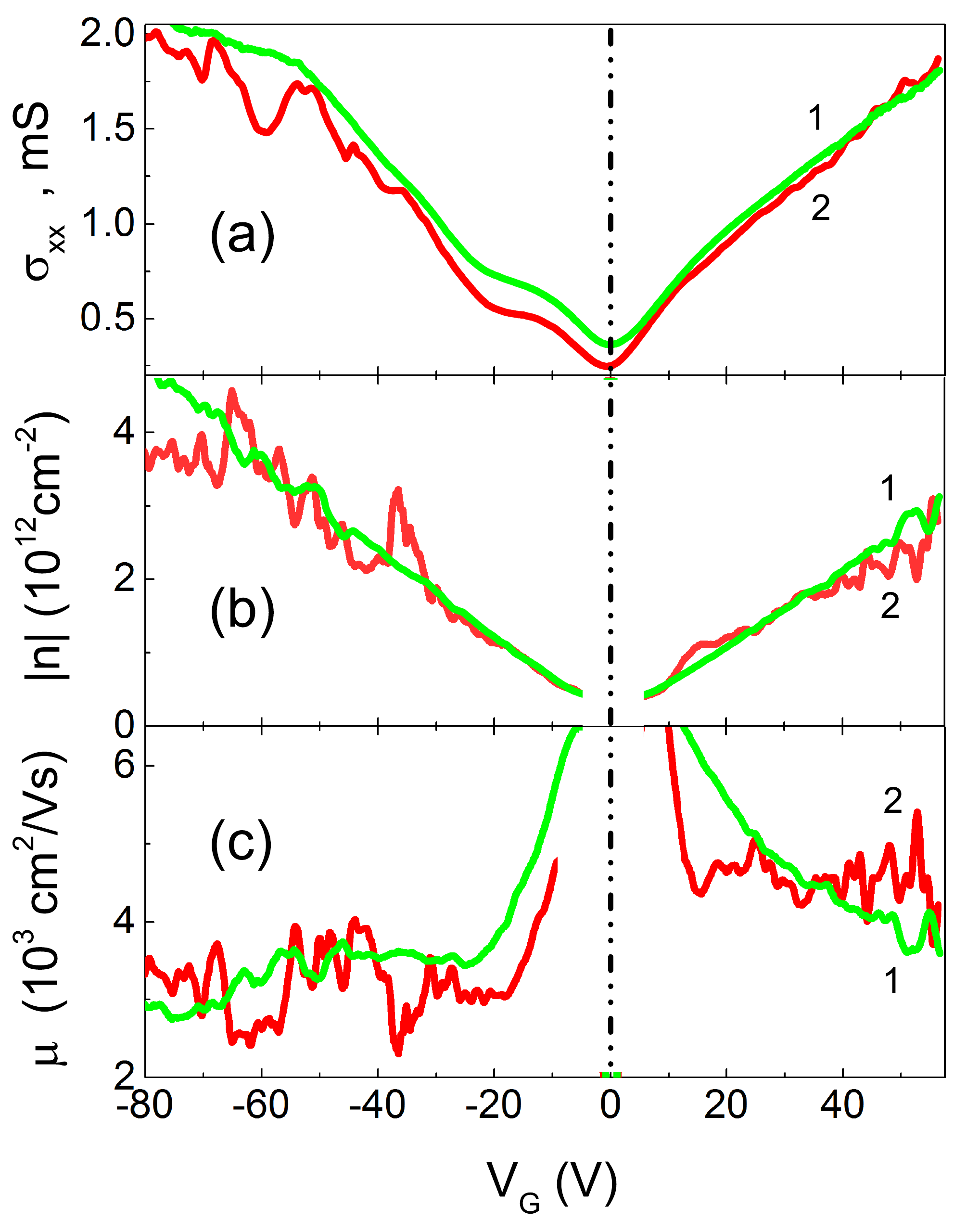}
	\caption{(Color online) Dependences of $\sigma_{xx}$, $|n|$ and $\mu$ plotted as a function of $V_{G}$ at $T = 50$ K (1) and 2 K (2).}
	\label{Layout4}
\end{figure}

 c) \underline{Mobility.}\\
 
 Figure 5 shows mobility as a function of $|n|$ for electrons and holes measured at different $T$. One sees that at low carrier density, mobility is asymmetric: $\mu_{n}> \mu_{p}$, which determines also the conduction asymmetry shown in Fig. 2. Previously, electron-hole conduction asymmetry was observed in the graphene FET geometry (three electrodes – source, drain and gate) \cite{18,19} and was explained by the imbalanced charge injection from the source and drain metal electrodes. In \cite{20}, it was shown that no significant difference was observed in the conduction asymmetry in devices with different metallic electrodes. Moreover, in our samples, the influence of the contact resistance is negligible due to the Hall bar sample geometry where potential probes are removed from the source and drain contacts. We believe that our results are in agreement with the model proposed in \cite{21}, where asymmetry of mobility in monolayer graphene is explained by the difference in the relativistic scattering cross-section, according to which the massless Dirac fermions are scattered more strongly when they are attracted to a charged impurity than when they are repelled from it. In our samples, the negatively charged acceptors determine the positive value of $V_{NP}$. As a result, positively charged holes are scattered by acceptors more strongly than electrons which explains the asymmetry in mobility and conductivity. \par

Figure 5 also shows that upon increasing of the carrier concentration, the asymmetry decreases. This can be interpreted as due to the effective screening of the long-range charged impurity potential by mobile carriers and domination of the short-range scattering by neutral defects located in SiO$_{2}$ and in the graphene itself. The Boltzmann theory gives the following expression for the conductivity of massless Dirac fermions \cite{7}:

\begin{center}
$\sigma = \frac{e^{2}v^{2}_{F}}{2}D(E_{F})\tau(E_{F})$ (Eq. 3)
\end{center}
Where $D(E_{F}) = 2E_{F}/\pi (\hbar v_{F})^{2} = \hbar v_{F} n^{1/2}$ is the density-of-states in monolayer graphene, $\tau (E_{F})$ is the relaxation time at the Fermi level.  This gives for mobility 
\begin{center}
$\mu = \frac{ev^{2}_{F}}{E_{F}}\tau (E_{F}) \propto \frac{1}{\sqrt[•]{n}}\tau(E_{F})$ (Eq. 4)
\end{center}

Theoretical consideration \cite{7} shows that for the long-range scattering by charged impurities, $\tau(E_{F}) \sim n^{1/2}$, while for the short-range scattering by vacancies, dislocations and other neutral defects, $\tau(E_{F}) \sim 1/n^{1/2}$. As a result, one gets from Eq. (4) that when the long-range scattering by charged impurities dominates, $\mu \approx$ const and does not depend on the carrier density, while domination of the short-range scattering by neutral centers leads to $\mu \sim 1/n$ which results in a negative correlation between mobility and carrier concentration \cite{22}. The dependence of the hole mobility in Fig. 5 clearly demonstrates the transition from long-range to short-range scattering at $p \approx 3\times 10^{12}$ cm$^{-2}$.\par 

The temperature dependence of mobility is different: at high temperatures, mobility increases with decreasing $T$, but become temperature-independent below 100 K. This shows that at high temperatures, scattering of carriers on phonons dominates, but below 100 K phonon scattering is insignificant which results in temperature-independent mobility.\\

\begin{figure}[H]
	\centering
	\includegraphics[scale=0.33]{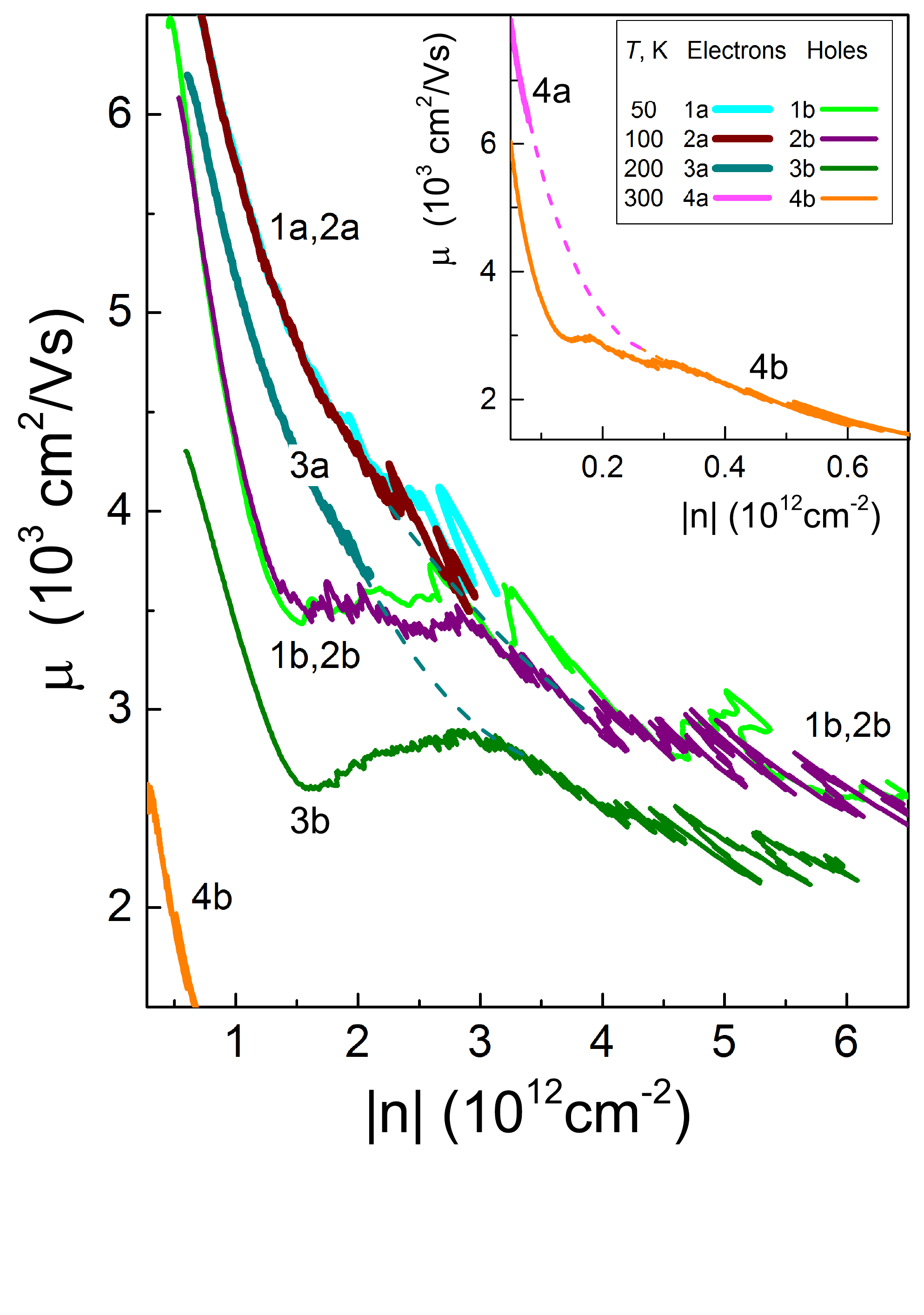}
	\caption{(Color online) Mobility of electrons (a) and holes (b) plotted as a function of $|n|$ at different temperatures.  $T$: 1- 50 K, 2 – 100 K, 3 – 200 K and 4 – 300 K. Mobility at 300 K is shown separately in insert. The dashed lines are guides to the eyes.}
	\label{Layout5}
\end{figure}

d) \underline{Quantum Hall Effect.}\\

Figure 6 shows that in strong perpendicular magnetic fields $B > 6$ T, Shubnikov-de Haas (SdH) oscillations of conductivity are observed together with the half-integer quantum Hall plateaus. The latter are characteristic only for monolayer graphene with massless Dirac fermions as charge carriers \cite{1,23}. In Fig. 6a, the reversal resistivity $(\rho_{xx})^{-1}$ is plotted in a weak and strong magnetic fields. However, in strong fields, when the perpendicular component of resistivity $\rho_{xy}$ is not negligible, one have to take into account that the relation between $\sigma_{xx}$ and $\rho_{xx}$ in two-dimensions has the form

\begin{center}
$\sigma_{xx}=\frac{\rho_{xx}}{\rho^{2}_{xx}+\rho^{2}_{xy}}$ (Eq. 5)
\end{center}

The dependence of $\sigma_{xx}(V_{G})$, which is plotted in Fig. 6b, clearly shows the SdH oscillations in the electron and hole branches of the conductivity. The positions of $V_{G}$ which correspond to the SdH minimum allow us to estimate the concentration of electrons needed to fill the Landau sub-bands.\par

\begin{figure}[H]
	\centering
	\includegraphics[scale=0.33]{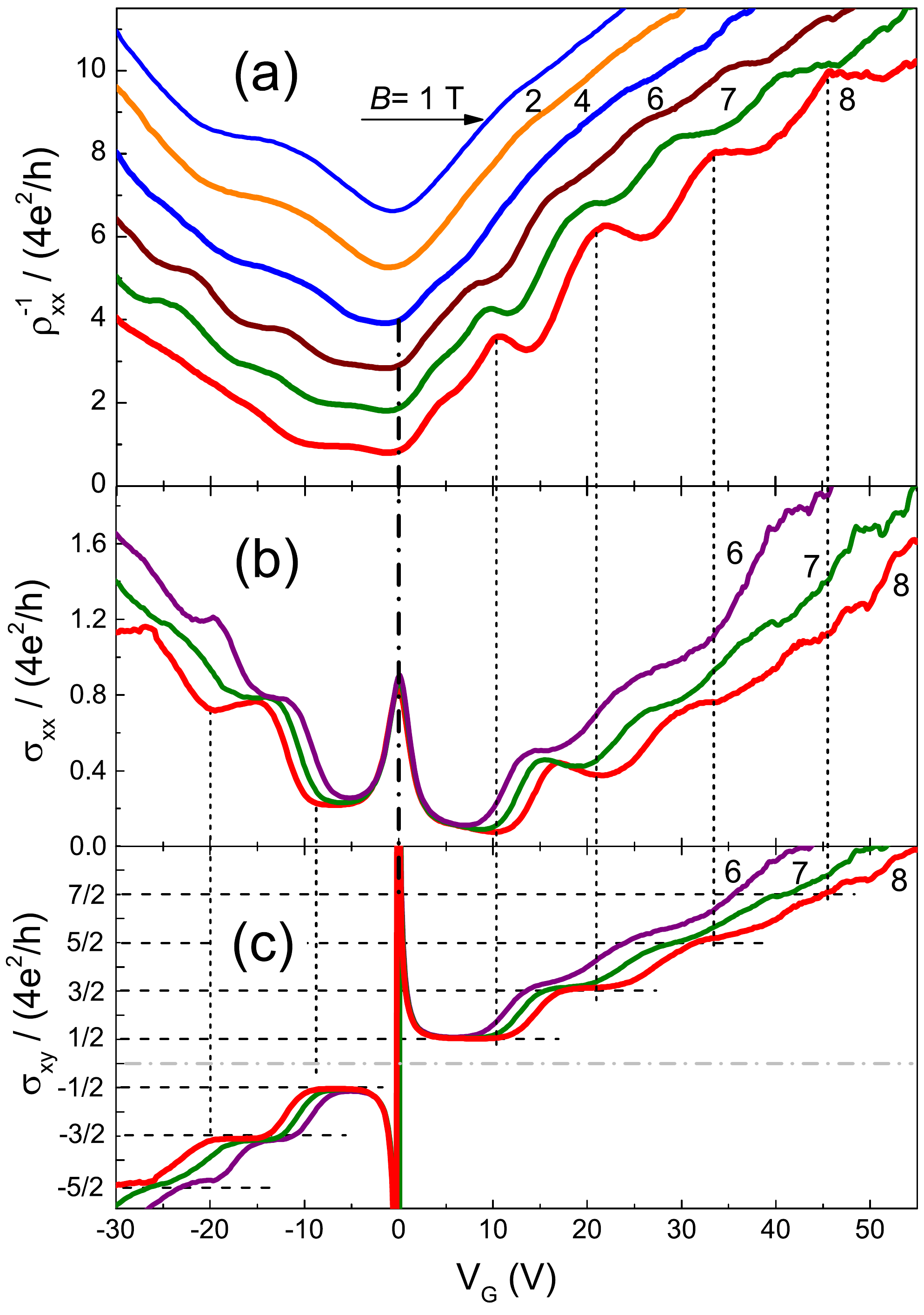}
	\caption{(Color online) (a) $\sigma_{xx}=(\rho_{xx})^{-1}$, (b) two-dimensional conductivity $\sigma_{xx}=\rho_{xx}/(\rho^{2}_{xx}+\rho^{2}_{xy})$, (c) $\sigma_{xy}$ in different magnetic fields, plotted as a function of $V_{G}$ in units $(4e^{2}/h)$. Curves in (a) are shifted by one for clarity. Numbers near curves indicate $B(T)$. Dotted vertical lines correspond to SdH minimums at $B = 8$ T, $T = 2$ K.}
	\label{Layout6}
\end{figure}

The energy of Dirac fermions in a magnetic field $B$ is proportional to $(B)^{1/2}$. For the filled $i$ and $i+1$ Landau sub-bands ($i$ is the integer 1, 2, 3, …), one can write the following relationship for electrons at the Fermi energy:

\begin{center}
$E_{F}=\hbar \omega^{i}_{c}\sqrt[•]{i}$, $E_{F}=\hbar \omega^{i+1}_{c}\sqrt[•]{i+1}$ (Eq. 6)
\end{center}

Here the "cyclotron frequency", $\omega^{i}_{c} = 2^{1/2}(v_{F}/\lambda) = v_{F}(2eB_{i}/\hbar)^{1/2}$, magnetic length $\lambda = (\hbar/eB_{i})^{1/2}$, the Fermi velocity $v_{F}  = 10^{6}$ m/s, and $B_{i}$ is the magnetic field which corresponds to the $i$-th Landau sub-band. Without a magnetic field, the density of electrons $n$ determines the position of Fermi energy $E_{F}: n = g_{s}g_{v}E_{F}^{2}/4\pi\hbar^{2}v_{F}^{2}$, where $g_{s} = 2$ is the spin degeneracy and $g_{v} = 2$ is the valley degeneracy. This gives

\begin{center}
$\pi\hbar^{2} n_{i}=E_{F}^{2}/v_{F}^{2}=2ieB_{i}\hbar$ (Eq. 7)
\end{center}

When the magnetic field $B$ is fixed, one can determine the carrier concentration $n_{i}$ to fill $i$ sub-bands

\begin{center}
$n_{i}=i(2eB/\pi\hbar)\approx i10^{11}B$ cm$^{-2}$T$^{-1}$ (Eq. 8)
\end{center}

Using Eq. (8), we obtain that for $B = 8$ T, $n_{i}$ is multiple to $8\times 10^{11}$ cm$^{-2}$ i.e. $n_{i} = 0.8, 1.6, 2.4$ and $3.2\times 10^{12}$ cm$^{-2}$ for $i = 1, 2, 3, 4$. One can see from Fig. 6 that for electrons, the SdH minimums at $B = 8$ T, occur at $V_{G} =$ 10.3, 21.0, 33.4, and 45.7 V. These four points are shown in Fig. 3 as open circles. Correspondingly, the points for $B = 7$ T were also calculated in this manner and shown in Fig. 3 as triangles. Similar calculations were made for SdH minimums on the $p$-side of $\sigma_{xx}(V_{G})$ curves and corresponding points are also shown in Fig. 3. One sees in Fig. 3 that all calculated points coincide with the values of the total density of electrons and holes induced by the oxide capacitance $C_{ox}$. This shows that in strong magnetic fields, all charge carriers participate in the filling of Landau sub-bands, as opposed to the Hall effect.

\section{conclusion}

The following new results were obtained by investigating the electron transport in large scale commercial CVD-grown polycrystalline monolayer graphene samples:\\

1)	Temperature dependence of conductivity as a function of the gate voltage $\sigma(V_{g})$ showed the following features: position of $V_{g}$ at the charge neutrality point, $V_{NP}$, is positive and large at 300 K, it first decreases with decreasing $T$ and then saturates and remains positive below 100 K. The width of hysteresis also decreases and then disappears at low $T$. The positive value of $V_{NP}$ is usually explained by unintentional doping of graphene by holes due to the existence of negatively charged acceptors. Therefore, decreasing $V_{NP}$ with decreasing $T$ can be explained by the freeze-out of holes on acceptors.  The absence of the shift below 100 K shows that some acceptors are located in SiO$_{2}$ substrate near the interface. This results in separation of holes from these acceptors by potential barrier and to a significant increase of the freeze-out time at low $T$. The latter also explains the disappearance of the hysteresis which is usually associated with charging and recharging of the impurity centers. \\

2)	At low carrier density $|n|$, significant asymmetry is observed: mobility of holes is less than mobility of electrons. With increasing $|n|$, electron mobility decreases and asymmetry disappears.  Asymmetry is explained by the difference in the relativistic scattering of massless Dirac fermions by negatively charged acceptors in attractive or repulsive potential. Disappearance of asymmetry with increasing $|n|$ is explained by the screening of the long-range charged impurity potential and prevalence of the short-range scattering by neutral defects.\\

3)	It was found that the value of $|n|$ determined from the Hall effect is less than the value induced by the gate voltage. This can be explained by the assumption that the charge carriers in polycrystalline samples have different mobility at different places (for example, near the grain boundaries) which leads to reduction of $|n|$ determined from the Hall effect measurements. It was found, however, that the values of $|n|$ determined from the Shubnikov-de Haas (SdH) oscillations in strong perpendicular magnetic fields are equal to the full values of carriers induced by the gate voltage. This shows that all charge carriers participate in the filling of Landau sub-bands, as opposed to the Hall effect.


\begin{thebibliography}{2}
	\bibitem {1} K.S. Novoselov, A. K. Geim, S. V. Morozov, D. Jiang, M. I. Katsnelson, I. V. Grigorieva, S. V. Dubonos, and A. A. Firsov,  Nature \textbf{438}, 197 (2005).
	\bibitem {2} V.E. Dorgan, M.-H. Bae, and E. Pop, Appl. Phys. Lett. \textbf{97}, 082112 (2010).
	\bibitem {3} K.I.  Bolotin, K. J. Sikes, Z. Jiang, M. Klima, G. Fudenberg, J. Hone, P. Kim, and H. L. Stormer, Solid State Commun. \textbf{146}, 351 (2008).
	\bibitem {4} S.V. Morozov,  K. S. Novoselov, M. I. Katsnelson, F. Schedin, D. C. Elias, J. A. Jaszczak, and A. K. Geim., Phys. Rev. Lett. \textbf{100}, 016602 (2008).
	\bibitem {5} J. Martin N. Akerman, G. Ulbricht, T. Lohmann, J. H. Smet, K. von Klitzing, and A. Yacoby, Nature Physics \textbf{4}, 144 (2008).
	\bibitem {6} A.K. Geim, and K.S. Novoselov, Nature Materials \textbf{6}, 183 (2007).
	\bibitem {7} S. Das Sarma, S. Adam, E. H. Hwang, and E. Rossi, Rev. Mod. Phys. \textbf{83}, 407 (2011).
	\bibitem {8} I. Shlimak, A. Haran, E. Zion, T. Havdala, Y. Kaganovskii, A.V. Butenko, L. Wolfson, V. Richter, D. Naveh, A. Sharoni, E. Kogan, and M. Kaveh, Phys. Rev. B \textbf{91}, 045414 (2015).
	\bibitem {9} E. Zion, A. Haran, A.V. Butenko, L. Wolfson, Y. Kaganovskii, T. Havdala, A. Sharoni, D. Naveh, V. Richter, M. Kaveh, E. Kogan, and I. Shlimak, Graphene \textbf{4}, 45 (2015).
	\bibitem {10} I. Shlimak, E. Zion, A.V. Butenko, L. Wolfson, V. Richter, Y. Kaganovskii, 
A. Sharoni, A. Haran, D. Naveh, E. Kogan, and M. Kaveh, Physica E \textbf{76}, 158 (2016).
	\bibitem {11} A. Butenko, E. Zion, Y. Kaganovskii,  L. Wolfson, V. Richter,  A. Sharoni, E. Kogan, M. Kaveh, and I. Shlimak, J. Appl. Phys. \textbf{120}, 044306 (2016). 
	\bibitem {12} E. Zion, A. Butenko, Y. Kaganovskii, V. Richter,  L. Wolfson,  A. Sharoni, E. Kogan, M. Kaveh, and I. Shlimak, J. Appl. Phys/ \textbf{121}, 114301 (2017).
	\bibitem {13} H. Wang, Y. Wu, C. Cong, J. Shang, and T. Yu, ACS Nano \textbf{4}, 7221 (2010).
	\bibitem {14} G. Bohra, R. Somphonsane, N. Aoki, Y. Ochiai, D. K. Ferry, and J. P. Bird, Appl. Phys. Lett. \textbf{101}, 093110 (2012). 
	\bibitem {15} B. Liu, R. Akis, D.K. Ferry, G. Bohra, R. Somphonsane, H. Ramamoorthy, and J. P. Bird, J. Phys.: Condens. Matter \textbf{28}, 135302 (2016).
	\bibitem {16} W. Zhu, V. Perebeinos, M. Freitag, and P. Avouris, Phys. Rev. B \textbf{80}, 235402 (2009).
	\bibitem {17} G.I. Zebrev, arXiv:1102.2348v1.
	\bibitem {18} S. Barraza-Lopez, M. Vanević, M. Kindermann, and M. Y. Chou, Phys. Rev. Lett. \textbf{104}, 076807 (2010).
	\bibitem {19} W.-R. Hannes, M. Jonson, and M. Titov, Phys. Rev. B \textbf{84}, 045414 (2011).
	\bibitem {20} P.K. Srivastava, S. Arya, S. Kumar, and S. Ghosh, arXiv:1608.05808.
	\bibitem {21} D.S. Novikov, Appl. Phys. Lett. \textbf{91}, 102102 (2007).
	\bibitem {22} J. Lu, J. Pan, S.-S. Yeh, H. Zhang, Y. Zheng, Q. Chen, Z. Wang, B. Zhang, J.-J. Lin, and P. Sheng, Phys. Rev. B \textbf{90}, 085434 (2014).
	\bibitem {23} Y. Zhang, Y.-W. Tan, H. L. Stormer, and P. Kim, Nature \textbf{438}, 201 (2005).				
	
\end{thebibliography}
\end{document}